\documentclass[11pt]{article}

\usepackage{graphicx}

\usepackage{bm}

\begin{document}


\title{C-axis critical current of a PrFeAsO$_{0.7}$ single crystal} 

\author{
\textsc{H. Kashiwaya},$^{1}$ 
\textsc{K. Shirai},$^{1,2}$ 
\textsc{T. Matsumoto},$^{1}$
\textsc{H. Shibata},$^{1}$ \\
\textsc{H. Kambara},$^{1}$ 
\textsc{M. Ishikado},$^{3,4}$ 
\textsc{H. Eisaki},$^{1,4}$ 
\textsc{A. Iyo},$^{1,4}$ \\
\textsc{S. Shamoto},$^{4}$
\textsc{I. Kurosawa},$^{2}$ 
and \textsc{S. Kashiwaya}$^{1}$}
\date{}
\maketitle


$^{1}$National Institute of Advanced Industrial Science and Technology (AIST), Ibaraki, 305-8568, Japan\\
$^{2}$Japan Women's University, Tokyo 112-8681, Japan \\
$^{3}$Japan Atomic Energy Agency, Ibaraki 319-1195, Japan\\
$^{4}$JST, Transformative Research-Project on Iron Pnictides (TRIP), Tokyo 102-0075, Japan \\

\begin{abstract}
The $c$-axis transport properties of a high-pressure synthesized PrFeAsO$_{0.7}$ single crystal are studied using s-shaped junctions.
Resistivity anisotropy of about 120 detected at 50 K shows the presence of strong anisotropy in the electronic states.
The obtained critical current density for the $c$-axis of 2.9$\times$10$^5$ A/cm$^2$ is two orders of magnitude larger than that in Bi$_2$Sr$_{1.6}$La$_{0.4}$CuO$_{6+\delta}$.
The appearance of a hysteresis in the current-voltage curve below $T_c$ is the manifestation of the intrinsic Josephson effect similar to that in cuprate superconductors.
The suppression of the critical current-normal resistance ($I_cR_n$) product is explained by an inspecular transport in s$_\pm$-wave pair potential.
\end{abstract}



The discovery of a family of iron-pnictide superconductors has renewed our interests in unconventional superconductors.\cite{Kamihara} 
The stack of superconducting FeAs sheets sandwiched between blocking layers characterizes the crystal structures of the iron-pnictides.
The similarity of the crystal structures to those of the cuprate superconductors suggests the realization of strong anisotropic electronic states. 
In contrast, previously reported experimental data on (Ba,K)Fe$_2$As$_2$ ($T_c$ $\sim$ 28 K) have suggested nearly isotropic features in the temperature range between 10-27 K based on $H_{c2}$ measurements. \cite{Yan}
Since anisotropy has been the bottleneck in several possible applications, such as power supply cables, the exact evaluation of the $c$-axis transports on iron-pnictides is an important issue.
\par
Here, we present the $c$-axis transport properties and the anisotropy of an oxygen deficient PrFeAsO$_{0.7}$ single crystal evaluated using s-shaped junctions fabricated by a focused ion beam (FIB) process.
PrFeAsO$_{0.7}$ is one of the LnFeAsO (Ln=lanthanide, so-called $\lq$1111$\lq$) compounds having relatively higher anisotropy among the iron-pnictides.\cite{Kito,Miyazawa,Ishikado}
We also applied the same measurements to Bi$_2$Sr$_{1.6}$La$_{0.4}$CuO$_{6+\delta}$ (Bi2201) single crystals as a reference.
Both compounds are single layer systems with similar $Tc$'s, which makes it easier to clarify the differences between iron-pnictides and cuprates.
The main differences between the two are the pair potential symmetry and the band structure as depicted in Table I.
In the case of cuprates, the $c$-axis transport below $T_c$ is dominated by the interlayer Josephson effect, so-called intrinsic Josephson effect, that has been identified by a large hysteresis in the current-voltage ($I$-$V$) curve.\cite{Kleiner}
The applications of the intrinsic Josephson junction (IJJ) include a terahertz radiation source and a qubit.
Therefore, one aspect of our investigation is whether a similar Josephson effect can be observed in the iron-pnictides.
\par
The single crystals of oxygen deficient PrFeAsO$_{0.7}$ and Bi2201 were prepared by a high-pressure synthesis method using belt-type anvil apparatus and by a floating zone method, respectively. 
The crystals were fixed on SrTiO$_3$ substrates after they were cut into pellets with a size of 10-100$\mu$m.
Then the center parts of the crystals were necked down to 2-3 $\mu$m from the top using a FIB.
The $ab$-plane resistivity $\rho_{ab}(T)$ was measured in this configuration.
The necked devices were processed further by a FIB radiated from the horizontal direction to form two slits.
The slits were designed to have an overlap along the $c$-axis so that the current direction was restricted to the $c$-axis in the necked region.
Typical scanning ion microscopy images of s-shaped junctions are shown in Fig. 1.
The junction sizes of 1-2$\mu$m were small enough to be regarded as short junctions.
The present device configuration has widely been used for the IJJ in recent experiments.\cite{Latyshev}
Details of the crystal growth condition and the device fabrication process have been described elsewhere.\cite{Kito,Miyazawa,Ishikado,Shirai}
It should be noted that one of the essential advantages of the present device is that the influence of surface or interface degradation can be completely eliminated, because the present junction does not rely on the hetero-structure.
In addition, junction size of a few micrometers is small enough to exclude the unanticipated presence of grain boundaries inside the junction.
Thus the present method permits the unambiguous detection of the intrinsic crystal nature.
\par
Figure 1 shows the temperature dependences of resistivity for the $c$-axis $\rho_c(T)$ and the resistivity anisotropy $\gamma_\rho(T)$ determined by the ratio of $\rho_c(T)$ to $\rho_{ab}(T)$.
The resistance was measured with an ac current modulation of about 10$\mu$A.
In the case of Bi2201, $\rho_c(T)$ below 140K is insulating whereas that above 140K is metallic. 
A similar feature has been detected widely in various cuprates.
In contrast, $\rho_c(T)$ for PrFeAsO$_{0.7}$ is insulating for the entire temperature range.
The variation of resistivity of less than 10$\%$ across the temperature range from 50K to 300K is far smaller than that of Bi2201.
For both compounds, values of $\gamma_\rho(T)$ in Fig.1 show a monotonic increment with lowering temperature.
The $\gamma_\rho(T)$ of about 120 at 50K is far larger than that detected in (Ba,K)Fe$_2$As$_2$,\cite{Yan} compatible with those of the 1111 compounds,\cite{Jia,Jaroszynski,Tanatar,Balicas} and far smaller than those of Bi-based cuprates.\cite{Kleiner}
This fact implies that the block layer in PrFeAsO$_{0.7}$ works as an insulating barrier although the barrier height is relatively low as compared to Bi-based cuprates.
\par
Figure 2 shows the temperature dependences of the critical current $I_c$($T$) obtained below $T_c$.
The detected Josephson currents in both compounds monotonically increase with lowering temperature.
The temperature dependences mostly follow Ambegaokar-Baratoff (AB) formula shown as solid lines.\cite{AB}
For more detailed comparison, we need fittings by taking account of the probability distribution of the switching current.\cite{Krasnov,HKashiwaya}
The critical current density for the $c$-axis direction $J_c$($T$) of 2.9$\times$10$^5$ A/cm$^2$ in PrFeAsO$_{0.7}$ is two orders of magnitude larger than that of Bi2201.
Assuming that $J_c$($T$) for the $ab$-plane is given by the product of the $c$-axis $J_c$($T$) and $\sqrt{\gamma_\rho}$, $J_c$($T$) of several MA/cm$^2$ could be attainable at 4.2K.
This value is comparable to that obtained in Ba(Fe$_{1-x}$Co$_x$)$_2$As$_2$ thin films.\cite{Film}
\begin{figure}
\includegraphics[height=7cm]{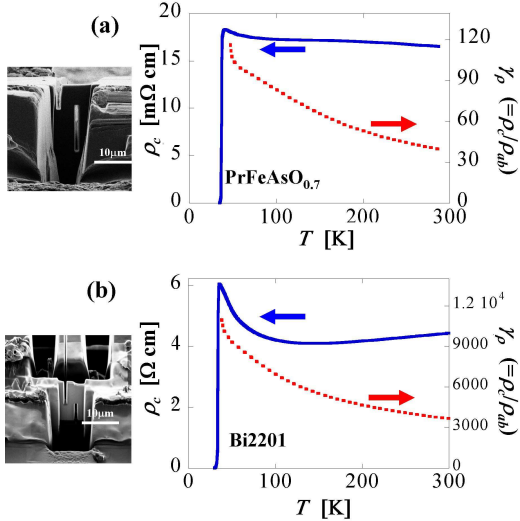}
\caption{ (color-online) Temperature dependences of $\rho_c$ and $\gamma_\rho$ for (a) PrFeAsO$_{0.7}$ and (b) Bi$_2$Sr$_{1.6}$La$_{0.4}$CuO$_{6+\delta}$. 
The scanning ion microscopy images of the s-shaped junctions used for the measurement are also shown.}
\end{figure}
\par
$I$-$V$ curves in the inset of Fig. 2 show the appearance of Josephson switching and the hysteresis for both Bi2201 and PrFeAsO$_{0.7}$.
Josephson switching means the discontinuous transition from the zero-voltage state to the finite voltage quasiparticle branch as the bias current increases.
We can evaluate damping of the junction from the switching dynamics.
The Q values estimated from the ratio of the switching and the retrapping current\cite{Tinkham} for Bi2201 and PrFeAsO$_{0.7}$ are 50 and 2 at 4.2K.
The low Q value in PrFeAsO$_{0.7}$ can be attributed to the low barrier height of the block layer and is consistent with the weakly insulating c-axis transport shown in Fig. 1.
An important question is whether the Josephson effect arises from the interlayer tunneling between adjacent FeAs layers similar to the intrinsic Josephson effect in cuprates.\cite{Kleiner}
We believe this interpretation is true for PrFeAsO$_{0.7}$ based on the reasons described below.
Firstly, the normal resistance ($R_n$) of the Josephson junction deduced from the gradient of the quasiparticle branch in the $I$-$V$ curve is about 10m$\Omega$.
In contrast, the transport measurement just above $T_c$ indicates that the resistance per one layer is about 30m$\Omega$ assuming that the s-shaped junction contains 1600 FeAs layers.
Since these two values are comparable, the origin of resistance in the Josephson junction is reasonably ascribed to the interlayer transfer.
This fact supports the intrinsic Josephson effect picture.
Secondly, we observed the appearance of the multiple branch structure by increasing the bias current.
The structure reflects the stacking of the Josephson junction in the $c$-axis direction, which is one of the manifestations of the intrinsic Josephson effect.\cite{Kleiner}
\begin{figure}
\includegraphics[height=7cm]{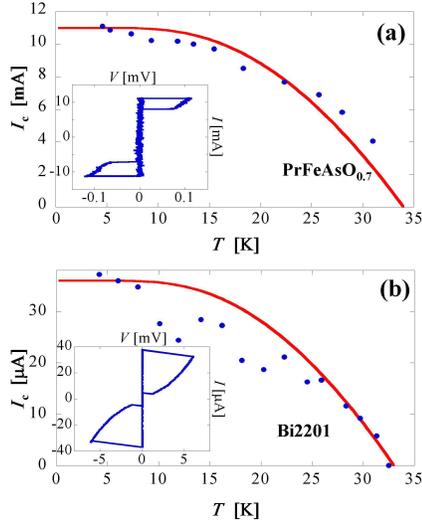}
\caption{  (color-online) Temperature dependences of $I_c$ for (a) PrFeAsO$_{0.7}$ (b) Bi$_2$Sr$_{1.6}$La$_{0.4}$CuO$_{6+\delta}$.
Solid lines represent $I_c$ based on the AB formula.
The insets show the typical $I$-$V$ curves obtained at 4.2K.}
\end{figure}
\par
Table I summarizes the data obtained in the present measurements.
One important difference between Bi2201 and PrFeAsO$_{0.7}$ is the $I_cR_n$ product.
In the case of Bi2201, the gap amplitude of 10-18mV has been obtained by scanning tunneling spectroscopy on the low temperature cleaved surfaces.\cite{Sugimoto}
This value corresponds not to the quasiparticle gap (40-100mV) but to the kink inside the quasiparticle gap.\cite{Alldredge}
The $I_cR_n$ product of 6mV estimated from the $I$-$V$ curve is comparable to the gap amplitude.
In contrast, the $I_cR_n$ product of 0.125mV in PrFeAsO$_{0.7}$ is two orders of magnitude smaller than the gap amplitude of 13.3mV detected by Andreev spectroscopy.\cite{Chen}
Following conventional theories of Josephson junctions that assume a simple barrier structure, such as no localized states inside the barrier, $I_cR_n$ at the zero point corresponds approximately to the gap amplitude both at the tunneling limit junction\cite{AB} and in the weak links.\cite{KO}
Therefore, such a small $I_cR_n$ cannot be attributed to the low barrier height of the block layer.
Another possibility is the suppression of the superconductivity near the junction interface.
Actually, the detection of the small $I_cR_n$ has been reported for the hetero-junctions of iron-pnictides\cite{Maryland}.
However, since the present result does not rely on the artificial interface or the cleaved surfaces, we can exclude this possibility.
\begin{table}
\caption{\label{tab:table1}Summary of experimental data for PrFeAsO$_{0.7}$ and Bi2201. 
The values in the upper columns have been obtained in the present experiment, and those in the lower columns are cited from references.
$J_c$, $Q$ and $I_cR_n$ are the values at 4.2K and $\gamma_\rho$ just above $T_c$.}
\begin{tabular}{lcc}
 & PrFeAsO$_{0.7}$ & Bi2201  \\
\hline
T$_c$[K] & 35 &  33\\
Anisotropy $\gamma_\rho$  & 120& $\sim$10000 \\
$dR/dT$ & Insulating & Insulating(T$<$140K)\\
C-axis $J_c$ [A/cm$^2$]& 2.9$\times$10$^5$ & 1000-2000\\
Q & 2 & 50 \\
$I_c R_n$[mV] & 0.125 & 6 \\
\hline
Gap amplitude [mV] & 13.3\cite{Chen} & 10-18(9K)\cite{Sugimoto} \\
Pair potential symmetry &$s_{\pm}$-wave & $d$-wave   \\
Band structure & multi-band & single band \\
\end{tabular}
\end{table}
\par
A plausible origin is the effect of the internal phase of s$_\pm$-wave symmetry.\cite{Mazin,Kuroki}
For an intuitive explanation, we assume a simplified superconductor having two isotropic pair potentials with a phase difference of $\pi$, $\Delta_{1}$ and -$\Delta_{2}$. 
In such case, $I_cR_n$ is roughly expressed by
$I_c R_n \propto\ \alpha \Delta_{1} + \beta \Delta_{2} - 2 \gamma \sqrt{\Delta_{1} \Delta_{2}}$,
where $\alpha$ and $\beta$ are parameters representing the Fermi surface information and the barrier height, and $\gamma$ is a parameter corresponding to the interband hopping due to the inspecularity ($\alpha$, $\beta$, $\gamma \geq 0$).
It is important to note that the minus in the equation comes from the phase difference of the two pair potentials.
For a system having complete translational symmetry for the $ab$-plane, the momentum in the plane is conserved through interlayer hopping.
$I_cR_n$ is approximately proportional to the amplitude of the pair potential integrated over the Fermi surface because $\gamma$ is zero even if the pair potential has anisotropy.\cite{SKashiwaya}
While in a real material, since the inspecular components inevitably exist, the deviation of $\gamma$ from zero reduces $I_c$.
The influence of such an effect is estimated to be small for the cuprates although it does exist.\cite{OKAnderson}
The present experimental result with PrFeAsO$_{0.7}$ implies that such an effect is far larger than that in cuprates, which results in the serious suppression of $I_cR_n$.
Y. Ota $et$ $al$. have discussed a similar effect at grain boundaries of an s$_\pm$-wave superconductor.\cite{Ota}
Since the present mechanism must be sensitive to the nature of the block layers, a systematic measurement for various iron-pnictides will reveal this effect more clearly.
\par
\par
We would like to thank Y. Yoshida, S. Kawabata and Y. Tanaka for fruitful discussion. 
This work was financially supported by Grant-in-Aid for Scientific Research (No.21710100, 20540392, 70393725) from JSPS, Japan and by Mitsubishi foundation.

%

\end{document}